\documentstyle[12pt]{article}

\newcommand{\beq}{\begin{equation}}
\newcommand{\eeq}[1]{\label{#1} \end{equation}}
\newcommand{\beqn}{\begin{eqnarray}}
\newcommand{\eeqn}{\end{eqnarray}}

\newcommand{\D}{{\cal{D}}}

\newcommand{\Z}{{\cal Z}}

\newcommand{\eq}[1]{eq.(\ref{#1})}

\begin{document}

\title{Explicit connection between conformal field theory and 2+1 Chern-Simons
theory}
\author{
Daniel C. Cabra$^a$\thanks{CONICET, Argentina. E-mail address:
cabra@venus.fisica.unlp.edu.ar} 
and  
Gerardo L. Rossini$^b$\thanks{On live of absence from
Universidad Nacional de La Plata and CONICET, Argentina.
E-mail address: rossini@ctpa03.mit.edu}
\\
~
\\
{\normalsize\it $^a$Departamento de F\'{\i}sica,}\\
{\normalsize\it Universidad Nacional de La Plata,}\\
{\normalsize\it C.C. 67 (1900) La Plata, Argentina}\\
~\\
{\normalsize\it
$^b$ Center for Theoretical Physics,}\\
{\normalsize\it Laboratory for Nuclear Science
and Department of Physics,}\\
{\normalsize\it Massachusetts Institute of Technology,}\\
{\normalsize\it Cambridge, MA 02139-4307, USA}\\ } 

\date{}

\maketitle

\begin{abstract}
{We give explicit field theoretical representations for the observables
of 2+1 dimensional Chern-Simons theory
in terms of gauge invariant composites of 2D WZW fields.
To test our identification we compute some basic Wilson loop
correlators reobtaining known results. }
\end{abstract}
\newpage

\pagenumbering{arabic}

Since the pionnering work of Witten \cite{W1}, where it was shown that the 
Hilbert space of pure Chern Simons (CS) theories is isomorphic to 
the space of conformal blocks of an underlying Conformal Field Theory (CFT), 
the connection between 1+1 conformal field theory and 2+1 CS
theory has been extensively studied \cite{2,G,W3,YS,BT}.

However, the relations between 1+1 CFT and CS related 2+1 phenomena seem to 
exceed the original setting of pure CS topological field theory. 
An still intriguing manifestation of this relation appears in the 
description of planar condensed matter systems, such as Quantum 
Hall Effect (QHE) systems \cite{LF}, where the CS action has found fruitful
application. In fact, QHE ground state wave functions have been constructed
using CFT data: in refs. \cite{MR,Wen,BW}, wave functions were constructed  
in terms of the conformal blocks of some CFT's 
such as the critical Ising model and $SU(N)_k$ Wess-Zumino-Witten models. 

In view of this, it is worthwhile to analize the 
connection from different approaches and trade the 
purely formal point of view of previous works for a more explicit and 
computationally workable approach. 
In particular, an explicit identification of
the observables in CS theory (Wilson loop operators) in terms of two
dimensional conformal fields would be useful to extend the connection to
more general cases than the pure CS one. This identification between the
observables at the operator level has  been only partially addressed in the
literature \cite{BT}. 

The purpose of the present paper is to study this point,
and to give a direct way to find the 
explicit expression of an arbitrary Wilson loop operator in terms 
of the corresponding two
dimensional conformal primary operators. 

To this end we recall
a recent version of the CS -- conformal field theory connection \cite{G}.
This approach uses the so-called transverse lattice
construction \cite{BPR} in 2+1 dimensions, linking 2-dimensional 
group $G$ level $k$ Wess-Zumino-Witten (WZW) models in each site through 
left-right asymmetric coupling with $G$ gauge fields.
In fact, it was
shown in ref.\cite{G} that the continuum limit of this construction
leads to a group $G$ pure CS theory.
Within this approach, Wilson loop correlators were evaluated in the
transverse lattice by using representation theory of chiral algebras in each 
WZW layer. Notice that one distinctive feature of this approach is that it 
does not undergo a dimensional reduction from the CS manifold to a 1+1 
dimensional manifold, but rather deals with a 2+1 dimensional array of CFTs.

Following this route we perform in this note the path integral quantization 
of the theory, giving explicit operator realizations
of the Wilson loop operators of the CS theory
in terms of gauge invariant composites of two dimensional WZW fields.
We show how to evaluate both the partition function and correlation 
functions for the observables by using decoupling techniques
in the path integral framework. We finally test our
construction with the computation of some basic Wilson loop correlators;
our results are of course in agreement with those originally presented by 
Witten \cite{W1}.

For the sake of clarity we first review the transverse lattice construction
in ref.\cite{G} and then present the operator realization
of the Wilson loop operators in this approach.

The transverse lattice geometry consists of a 2D manifold, which is taken
as a Minkowski space $M_2$, and a transverse discrete dimension,
which is taken as
a periodic chain of $N$ sites ($S^1$ topology). One introduces a 
field\footnote{Our conventions for light cone coordinates are
$x^{\pm}=\frac{1}{\sqrt 2}(x^0 \pm x^1)$.}
$g_n(x^+,x^-)$ on each site, governed by a level $k$ WZW action \cite{W2}
\beq
kW[g_n]=
\frac{k}{8\pi} \int_{M_2} d^2x
Tr(\partial_{\mu}g_n\partial^{\mu}g_n^{-1})
+ k\Gamma[g_n]
\eeq{1}
where $g_n$ takes values in a simple Lie group $G$ and $\Gamma[g_n]$
is the Wess-Zumino term
\beq
\Gamma[g]=\frac{1}{12\pi}\int_Y d^3y\epsilon^{ijk}
Tr(g^{-1}\partial_i g g^{-1}\partial_j g g^{-1}\partial_k g)
\eeq{2}
with $\partial Y =M_2$.
The coupling between two-dimensional layers is accomplished through
gauge fields $A_{\mu,n}$; $g_n$ is left-coupled to $A_{\mu,n}$ and
right-coupled to $A_{\mu,n+1}$. The corresponding interaction    term
in the action is given by \cite{GK}
\beqn
\lefteqn{I[g_n,A_{\pm,n},A_{\pm,n+1}]=\frac{k}{2\pi}
\int_{M_2}d^2x Tr[A_{-,n+1}g_n^{-1}\partial_+ g_n
-A_{+,n}\partial_- g_n g_n^{-1}}  \nonumber\\
& & + A_{+,n}g_nA_{-,n+1}g_n^{-1}
-\frac{1}{2}(A_{-,n}A_{+,n}+ A_{+,n+1}A_{-,n+1})].
\label{3}
\eeqn
The action $S_n=kW[g_n]+I[g_n,A_{\pm,n},A_{\pm,n+1}]$ is not invariant
under gauge transformations
\beqn
& \delta g_n(x^+,x^-)=
\Omega_n(x^+,x^-)g_n(x^+,x^-)-g_n(x^+,x^-)
\Omega_{n+1}(x^+,x^-),& \nonumber \\
& \delta A_{\pm,n}=\partial_{\pm} \Omega_n
+[\Omega_n,A_{\pm,n}],& \nonumber \\
&\delta A_{\pm,n+1}=\partial_{\pm} \Omega_{n+1}
+[\Omega_{n+1},A_{\pm,n+1}].&
\label{4}
\eeqn
Indeed, the change in the action $S_n$ reads
\beq
\delta S_n=\frac{k}{2\pi} \int_{M_2} d^2x Tr[
\Omega_n\epsilon^{\mu\nu}\partial_{\mu}A_{\nu,n}-
\Omega_{n+1}\epsilon^{\mu\nu}\partial_{\mu}A_{\nu,n+1}]
\eeq{5}
which is related to the non-Abelian anomaly in two dimensions \cite{PW}.
Note that in case the coupling was left-right symmetric ($A_{\pm,n}=
A_{\pm,n+1}$) the variation would vanish; this would be equivalent to gauging
the
anomaly free vector subgroup of the left and right global quiral symmetry of
the WZW model.

The entire system consists of a periodic chain of
2D layers. Its action is simply given by
\beq
S=\sum_{n=1}^{N} S_n.
\eeq{6}
This system is gauge invariant because of the cancellation of the second term
in (\ref{5}) with the first term coming from the variation corresponding
to the following site. We will refer to this interplay as gauge
invariance of the links (Fig.1).

In order to make contact with the 2+1 CS theory, following ref.\cite{G},
we represent the link field $g_n$ as a function on the transverse lattice,
\beq
g_n=\exp(-\int_{x^3}^{x^3+a} A_3(x^+,x^-,x^3) dx^3).
\eeq{7}
Here $a$ is the spacing of the lattice and $x^3=na$ will become a continuous
coordinate as $a\rightarrow 0$, $N\rightarrow\infty$, while $Na=L$ remains
constant. In this limit the phase exponent in (\ref{7}) can be written as
$-aA_3$, with $A_3$ evaluated in $x^3=(n+1/2)a$. Using this parametrization
in \eq{6} one readily obtains
\beq
\lim_{\begin{array}{cc} a\rightarrow 0\\ N \rightarrow \infty \end{array}}
S=\frac{k}{2\pi}\int_{M_2 \times S^1}d^3x \epsilon^{ijk}
Tr[A_i\partial_j A_k -\frac{2}{3} A_iA_jA_k] \equiv S_{CS},
\eeq{8}
which corresponds to the level $k$ CS action for the gauge
group $G$.
It should be stressed that the same result is obtained
starting from the
off critical WZW action instead of action (\ref{1}).

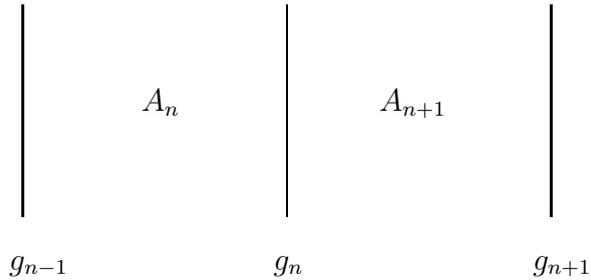
\begin{figure}
\begin{center}
\begin{picture}(200,150)(-100,-75)
\put(-100,50){\line(0,-1){80}}
\put(0,50){\line(0,-1){80}}
\put(100,50){\line(0,-1){80}}
\put(-5,-50){$g_n$}
\put(-105,-50){$g_{n-1}$}
\put(93,-50){$g_{n+1}$}
\put(-55,10){$A_n$}
\put(35,10){$A_{n+1}$}

\end{picture}

\caption{Gauge invariance of $S$ is achieved by compensation of gauge
anomalies at each link between WZW layers.}
\end{center}
\end{figure}

Once the connection (\ref{8}) between classical actions is established,
one is naturally
led to study the quantization of both theories and the relation
between their observables. This study has been performed in ref.\cite{G}
in the framework of canonical quantization, using the representation
theory of the Virasoro algebra for the 1+1-dimensional layers and then
solving the constraints arising from gauge invariance of $S$ in \eq{6}.
In this way, the correspondence between the physical states of the
lattice CS theory and some Wilson loops of the continuum CS
theory has been proved. In our investigation, we shall instead work in the
path-integral approach.
This will allow us to reach our aim, that is to construct explicit operator
expressions for Wilson loops in the lattice CS theory.

We start from the partition function of the
lattice CS theory
\beq
\Z=\int \prod_{n=1}^{N}\D A_{\mu,n} \D g_n e^{iS}
\eeq{9}
with $S$ given by \eq{6}. 

Before presenting our approach, we have to mention 
that the path integral (\ref{9}) can be effectively reduced to a 2-dimensional
theory\footnote{We thank G. Thompson for pointing this out to us}.
In fact, the results presented by Witten in ref.\cite{W3} show that 
one could shrink the transverse dimension to a point instead of taking the 
continuum limit (which is not surprising since being the pure CS theory 
a topological field theory, the length of the transverse dimension should be 
irrelevant). To be more precise, Witten has shown that
\beqn
\int & \D A_{\mu,n+1} \D g_n\D g_{n+1}&
e^{i(kW[g_n]+I[g_n,A_n,A_{n+1}] + 
kW[g_{n+1}]+I[g_{n+1},A_{n+1},A_{n+2}])}  =  \nonumber\\
& = & \int \D g e^{i(kW[g]+I[g_,A_n,A_{n+2}]}. 
\label{reduction}
\eeqn
This reduction formula eliminates the degrees of freedom associated to the 
site $n+1$, and its repeated application finally leads to the 
consideration of the 2-dimensional left-right symmetric $G/G$ coset model
\beq
\int \D A_{\mu} \D g e^{i(kW[g]+I[g,A,A])}.
\eeq{gg}
This model has been studied in the literature (see for instance \cite{YS,BT}).
In particular, in \cite{BT} the authors reobtain, as we will do, 
the already known results for Wilson loop expectation values in CS topological
field theory \cite{W1}. 
However, our aim 
is to give a way to construct operator realizations for Wilson loops 
in terms of conformal fields, and for this purpose the present approach 
is more appropriate, as we will show.

We then maintain every site in the chain and perform the following
(decoupling \cite{GK,Ka}) change of variables
\beqn
A_{+,n}=f_n^{-1}\partial_+f_n, \nonumber\\
A_{-,n}=h_n^{-1}\partial_-h_n, \nonumber\\
g_n=f_n^{-1}g_n^{(0)}h_{n+1}. \label{10}
\eeqn
After this change the action $S$ takes the form
\beq
S=\sum_{n=1}^{N}(kW[g_n^{(0)}] - kW[f_nh_n^{-1}]),
\eeq{11}
where the Polyakov-Wiegmann identity
\beq
W[gh]=W[g]+W[h]-\frac{1}{2\pi}\int d^2x Tr (g^{-1}\partial_+ g
\partial_-h h^{-1})
\eeq{ipw}
has been used \cite{IPW}.

The Jacobian associated with transformation (\ref{10}) is given by
\beq
\D A_{+,n} \D A_{-,n} = det(D_{+,n}^{Adj}) det(D_{-,n}^{Adj})
\D f_n \D h_n
\eeq{12}
where $D_{\pm,n}^{Adj}=\partial_{\pm} +[A_{\pm,n},\:\: ]$.
The adjoint determinants can be written as
\beq
det(D_{+,n}^{Adj}) det(D_{-,n}^{Adj})=\Z_{gh,n} e^{-2iC_v
W[f_nh_n^{-1}]}
\eeq{13}
where $C_v$ is the adjoint Casimir of $G$. Here we have adopted a
regularization prescription that preserves the gauge invariance 
(or anomaly cancellation) associated
to each link. The explicit form of the ghost partition function in
\eq{13} will not be relevant in the following.

As the last step to arrive at the decoupled partition function we change
$f_n \rightarrow \tilde{f}_n=f_nh_n^{-1}$ (with unit Jacobian). Then the
integral over $h_n$ factorizes as $Vol\hat{G}$. This procedure tantamounts
to fixing the $A_-=0$ gauge.
The decoupled partition function now reads
\beq
\Z=\Z_{gh} \int\prod_{n=1}^{N} \D g_n^{(0)} e^{ikW[g_n^{(0)}]}
\int \prod_{n=1}^{N} \D\tilde{f}_n
e^{-i(2C_v+k)W[\tilde{f}_n]}.
\eeq{14}

Thus the partition function factorizes in $3N$ sectors, corresponding to $N$
free ghost systems, $N$ $G$-valued WZW fields $g_n$ with level $k$ and
$N$ $G$-valued WZW fields $\tilde{f}_n$ with negative level $-(k+2C_v)$.
For each site $n$  one has a conformal field theory with vanishing
total central charge, built up from the different sectors:
a ghost sector contributing with $c_{gh}=-2dimG$, a level $k$ WZW
sector contributing with $c_k=k\, dimG/(k+C_v)$ and
a negative level WZW sector contributing with $c_{-(k+2C_v)}=
(k+2C_v)\, dimG / (k+C_v)$. The level $k$ field $g^{(0)}_n$ can be thought
as localized in the $n$-th site and the level $-(k+2C_v)$ field
$\tilde{f}_n$ as localized in the link between the ($n$-1)-th and the
$n$-th sites.
Notice that, although the partition function of the theory is completely
decoupled, BRST quantization condition connects the different sectors
(as will be apparent in \eq{18})
in order to
ensure unitarity \cite{Ka,KO}.

Once we have written the partition function in the simple form (\ref{14}),
we turn our attention to the construction of the observables of this
theory. We will then show that the lattice version observables
naturally lead to Wilson loop operators in the continuum CS limit.

The observables in the lattice are constructed from gauge invariant
composites of WZW fields, which in turn belong to integrable representations
of the group $G$. In order to keep a simple notation we will discuss
the specific example of $SU(2)_k$, although the extension to $SU(N)_k$ and
more general groups is straightforward.
The integrable representations of $SU(2)_k$ are characterized by the spin
$j=0,1/2,\dots, k/2$ \cite{GW}. The basic field $g_n$ in \eq{1} is taken in
the fundamental representation of $SU(2)$ ($j=1/2$). Fields $g_n^{(j)}$
in higher spin representations are constructed as appropriately symmetrized
direct products of fields in the fundamental one. As gauge invariance of the
lattice model is achieved in each link, in order to construct gauge invariant
composites one has to take the trace of the product of $g_n$ fields in each
site, all of them in the same representation. That is,
\beq
R_j(x^+,x^-)=Tr_j \prod_{n=1}^{N} g_n^{(j)}(x^+,x^-)
\eeq{15}
where $Tr_j$ means matrix trace in the representation of spin $j$.

To see the connection of these fields with Wilson loop operators
one has to use \eq{7} (now in the representation $j$) obtaining
\beq
R_j(x^+,x^-)=Tr_j \prod_{n=1}^{N} e^{-aA_3(x^+,x^-,na+a/2)}
\eeq{16}
which in the continuum limit gives
\beq
R_j(x^+,x^-) \longrightarrow_{a \rightarrow 0} Tr_j
P(e^{-\int_{C}dx^{\mu}A_{\mu}}).
\eeq{17}
This is the expression for Wilson loop operators winding once around a
circle $C$ passing through $(x^+,x^-)$ in each layer and carrying flux
in the representation $j$.

While the limit taken in (\ref{17}) is a standard procedure in lattice
gauge theories, our main result is that
the identification  is also valid at the level of quantum operators; in other 
words, one obtains the same result by taking the continuum limit 
before computing quantum expectation values, thus working in the CS theory, 
or after computing quantum expectation values in 2-dimensional conformal 
field theory.

We will check the previous statement by evaluating 
up to three point correlators
using the decoupled picture (\ref{14}). In this picture the fields in
\eq{15} can be written in terms of the decoupled variables $g_n^{(0)}$,
$\tilde{f}_n$ as
\beq
R_j(x^+,x^-) = Tr_j \prod_{n=1}^{N} (\tilde{f}_n^{(j)})^{-1}(x^+,x^-)
{g_n^{(0)}}^{(j)}(x^+,x^-) .
\eeq{18}
Thus, correlators involving $R_j$'s factorize in the level $k$ and level
$-(k+4)$ WZW sectors.

The conformal dimensions of the primary fields in a level $K$ $SU(2)$ WZW
theory are given by $C_j/(K+C_v)$, where $C_j=j(j+1)$ is the Casimir in the
spin $j$ representation and the adjoint Casimir is $C_v=2$.  For the fields
${g_n^{(0)}}^{(j)}$ and $\tilde{f}_n^{(j)}$ we have
\beqn
h[{g_n^{(0)}}^{(j)}]=\frac{j(j+1)}{k+2}\nonumber\\
h[\tilde{f}_n^{(j)}]=- \frac{j(j+1)}{k+2}\label{19}
\eeqn
and hence the conformal dimension of $R_j$ vanishes. This  implies that the
correlators are independent of the
coordinates
\footnote{This fact is a direct consequence of conformal symmetry for
the two and three point correlators.
In the four point case the analysis is more involved but coordinate
independence can be proved following ref.\cite{NS} (see section 3.2).}
$(x^+,x^-)$, this being in correspondence with the topological
nature  of Wilson loop operators in the pure CS theory.


The one point correlator vanishes except for the trivial $j=0$
representation, in which the fields correspond to the identity operator,
giving
\beq
\langle R_j \rangle = \delta_{j0}.
\eeq{20}

For the two point correlator one has to evaluate
\beq
\langle R_{j_1}(x_1) R_{j_2}(x_2)  \rangle =
\langle Tr_{j_1}( \prod_{n=1}^{N} g_n^{(j_1)}(x_1))
        Tr_{j_2}( \prod_{n=1}^{N} g_n^{(j_2)}(x_2))
          \rangle .
\eeq{2a}
In order to give a detailed derivation
we introduce indices for each representation $j$. As
$g^{(j)}$
transforms in the representation $j\times j$ of
$SU(2)_L \times SU(2)_R$, we write $g^{(j)}_{\alpha\alpha '}$ with
$\alpha (\alpha ') =-j, \dots ,j$ corresponding to the left (right)
representation. One can then factorize \eq{2a} as
\beqn
\lefteqn{ \prod_{n=1}^{N}\langle
({g_n^{(0)}}^{(j_1)})_{\alpha_n \alpha_n '} (x_1)
({g_n^{(0)}}^{(j_2)})_{\beta_n \beta_n '} (x_2)
 \rangle }
\nonumber \\
& & \prod_{n=1}^{N}
\langle ({\tilde{f}_n}^{(j_1)})^{-1}_{\alpha_{n-1}'\alpha_n}
(x_1)
({\tilde{f}_n}^{(j_2)})^{-1}_{ \beta_{n-1}' \beta_n}
(x_2)  \rangle
\label{2b}
\eeqn
where each v.e.v. in the first line of (\ref{23}) is evaluated in the theory
with action  $kW[g_n^{(0)}]$ and each one in the second line is evaluated
using the action $-(k+4)W[\tilde{f}_n]$.

It is known that the two point correlator of primary fields in 
a conformal field theory is
completely determined by conformal symmetry. For an $SU(2)$
WZW theory it reads \cite{FZ}
\beqn
\lefteqn{ \langle
({g}^{(j_1)})_{\alpha \alpha '} (x_1)
({g}^{(j_2)})_{\beta \beta '} (x_2)
 \rangle  = } \nonumber \\
& & (-1)^{2j_1-\alpha -\beta}\delta_{j_1j_2}\delta_{\alpha ,-\beta}
\delta_{\alpha ',-\beta '}(2x_{12}^+x_{12}^-)^{-2h_{j_1}},
\label{2c}
\eeqn
where $x_{12}=x_1-x_2$ and $h_j$ is the conformal dimension of the
fields in the representation $j$. Using this result in \eq{2b} one
readily obtains
\beq
\langle R_{j_1}(x_1) R_{j_2}(x_2) \rangle = \delta_{j_1 j_2}
\eeq{2d}

For the three point correlator the derivation is analogous.
One has to evaluate
\beqn
\lefteqn{\langle R_{j_1}(x_1) R_{j_2}(x_2) R_{j_3}(x_3) \rangle = }
\nonumber \\
& &\langle Tr_{j_1}( \prod_{n=1}^{N} g_n^{(j_1)}(x_1))
        Tr_{j_2}( \prod_{n=1}^{N} g_n^{(j_2)}(x_2))
        Tr_{j_3}( \prod_{n=1}^{N} g_n^{(j_3)}(x_3))  \rangle ,
\label{22}
\eeqn
which factorizes as
\beqn
\lefteqn{ \prod_{n=1}^{N}\langle
({g_n^{(0)}}^{(j_1)})_{\alpha_n \alpha_n '} (x_1)
({g_n^{(0)}}^{(j_2)})_{\beta_n \beta_n '} (x_2)
({g_n^{(0)}}^{(j_3)})_{\gamma_n \gamma_n '} (x_3) \rangle }
\nonumber \\
& & \prod_{n=1}^{N}
\langle ({\tilde{f}_n}^{(j_1)})^{-1}_{\alpha_{n-1}'\alpha_n}
(x_1)
({\tilde{f}_n}^{(j_2)})^{-1}_{ \beta_{n-1}' \beta_n}
(x_2)
({\tilde{f}_n}^{(j_3)})^{-1}_{\gamma_{n-1}'\gamma_n}
(x_3) \rangle .
\label{23}
\eeqn

The three point correlator of primary fields in a level $K$
conformal field theory is
determined by conformal symmetry only up to some coefficients which can
be evaluated from operator product expansion,
\beqn
\lefteqn{ \langle
({g}^{(j_1)})_{\alpha \alpha '} (x_1)
({g}^{(j_2)})_{\beta \beta '} (x_2)
({g}^{(j_3)})_{\gamma \gamma '} (x_3) \rangle  =
C\left(
        \begin{array}{ccc}
                j_1 & \alpha & \alpha ' \\
                j_2 & \beta  & \beta ' \\
                j_3 & \gamma & \gamma '
        \end{array}
                \right)_K \times}  \nonumber \\
& & (2x_{12}^+x_{12}^-)^{h_{j_3}-h_{j_1}-h_{j_2}}
(2x_{23}^+x_{23}^-)^{h_{j_1}-h_{j_2}-h_{j_3}}
(2x_{13}^+x_{13}^-)^{h_{j_2}-h_{j_1}-h_{j_3}},
\label{24}
\eeqn
where $x_{ij}=x_i-x_j$. For a level $K$ $SU(2)$ WZW theory the
coefficients $C$ have been given in ref.\cite{FZ} in the form
\beq
C\left( \begin{array}{ccc}
                j_1 & \alpha & \alpha ' \\
                j_2 & \beta  & \beta ' \\
                j_3 & \gamma & \gamma '
        \end{array}                             \right)_K  =
\left[  \begin{array}{ccc}
                j_1 & j_2 & j_3 \\
                \alpha & \beta & \gamma
        \end{array}                             \right]
\left[  \begin{array}{ccc}
                j_1 & j_2 & j_3 \\
                \alpha ' & \beta ' & \gamma '
        \end{array}                             \right]
\rho_K (j_1,j_2,j_3)
\eeq{25}
for $|j_1-j_2| \leq j_3 \leq min(j_1+j_2, k-j_1-j_2)$ and zero otherwise.
The first two factors in \eq{25} are the Wigner 3$j$-symbols and the
function $\rho_K (j_1,j_2,j_3)$ can be written in terms of $\Gamma$
functions; for our purposes it will be important that
\beq
\rho_k (j_1,j_2,j_3) \rho_{-(k+4)} (j_1,j_2,j_3) =1.
\eeq{26}
The exponents in \eq{24} are given in terms of the conformal weights of
the primary
fields; they cancel in our case in virtue of \eq{19}.

Using the results above the correlator in \eq{22} can be written as
\beqn
\lefteqn{ \langle R_{j_1}(x_1) R_{j_2}(x_2) R_{j_3}(x_3) \rangle =
\prod_{n=1}^{N}
C\left( \begin{array}{ccc}
                j_1 & \alpha_n & \alpha_n ' \\
                j_2 & \beta_n  & \beta_n ' \\
                j_3 & \gamma_n & \gamma_n '
        \end{array}                             \right)_k
C\left( \begin{array}{ccc}
                j_1 & \alpha_{n-1}' & \alpha_n  \\
                j_2 & \beta_{n-1}'  & \beta_n  \\
                j_3 & \gamma_{n-1}' & \gamma_n
        \end{array}                             \right)_{-(k+4)}   }
        \nonumber \\
& & = \left(
\left[  \begin{array}{ccc}
                j_1 & j_2 & j_3 \\
                \alpha_n & \beta_n & \gamma_n
        \end{array}                                   \right]
\left[  \begin{array}{ccc}
                j_1 & j_2 & j_3 \\
                \alpha_n & \beta_n & \gamma_n
        \end{array}                                   \right]
                                                                \right)^N
\left(
\left[  \begin{array}{ccc}
                j_1 & j_2 & j_3 \\
                \alpha_i ' & \beta_i ' & \gamma_i '
        \end{array}                                   \right]
\left[  \begin{array}{ccc}
                j_1 & j_2 & j_3 \\
                \alpha_i ' & \beta_i ' & \gamma_i '
       \end{array}                                   \right]
                                                                \right)^N
                                                                \nonumber\\
& &\label{27}
\eeqn
Orthogonality relations between the 3j-symbols allows one to express the
result as
\beq
\langle R_{j_1}(x_1) R_{j_2}(x_2) R_{j_3}(x_3) \rangle =
\delta(j_1,j_2,j_3)
\eeq{28}
where $\delta(j_1,j_2,j_3)$ means 1 in case that $j_1$, $j_2$ and $j_3$
satisfy a triangular condition and 0 otherwise.

Our results (\ref{20}), (\ref{2d}) and (\ref{28}) coincide with the
expectation value of one, two and three
unknotted Wilson loops given by Witten \cite{W1} in terms of the fusion rules
of the symmetry group, which for $SU(2)_k$ explicitly read \cite{Ver}
\beq
N^0_{j_1 j_2} = \delta_{j_1 j_2}
\eeq{29}

\beq
N_{j_1 j_2 j_3} = \left\{ \begin{array}{ll}
1 & \mbox{if } |j_1-j_2| \leq j_3 \leq min(j_1+j_2, k -j_1 -j_2)\\
0 & \mbox{otherwise.} \end{array} \right.
\eeq{30}
Let us recall that these correlators in the manifold $S^2\times S^1$ are
the basic results that, through the use of surgery techniques, allow for
the computation of the link invariants associated to arbitrary links in
$S^3$ \cite{W1}.

In summary, we have given an explicit operator realization of the
Wilson loop operators (winding once around $S^1$ and carrying
flux in the representation $j$) in terms of gauge invariant products of
WZW fields in the transverse lattice formulation. We have also shown
how to decouple the lattice partition function and in order to test this
scheme we have
explicitly evaluated correlators of up to three unknotted Wilson loops.
This results are not new, but they are of course in agreement with those 
found in the literature \cite{W1,BT}. 
However, our Lagrangian approach is new and complimentary to the direct 
3-dimensional CS approach and the 2-dimensional Hamiltonian approach, and 
could be interesting in regard to generalizations.
Following our presentation, one could for instance look
for explicit representations of Wilson loop operators in less trivial
topologies.

As stated in the introduction, the explicit identification between
observables in the CS theory and two dimensional primary fields could be
useful to exploit the connection in more general cases. In particular,
it is interesting to analize perturbations that violate the topological
symmetry of the CS theory but preserve the integrability of the
conformal theories in the
layers of the transverse lattice construction.
Another interesting open route arises from the fact that our
treatment of the underlying
conformal symmetry does not use representation theory of chiral algebras
but rather represents observables explicitly. It makes possible
to find connections with an alternative description of conformal field
theory, namely the spinon formulation \cite{Hal},\cite{L}.
These issues are under current investigation.

\vspace{5mm}
{\em Acknowledgments:} We wish to thank E.\ F.\ Moreno and
F.\ A.\ Schaposnik for helpful comments. This work was partially supported 
by CONICET, Argentina.

\end{document}